%% file: main.tex
\documentclass[conference]{IEEEtran}
\usepackage{cite}

\ifCLASSINFOpdf
\else
\fi
\usepackage{subfiles}
\usepackage{dirtytalk}
\usepackage{graphicx}
\usepackage{setspace}

\usepackage{hyperref}
\hypersetup{
    colorlinks = true,
    citecolor = {blue},
    urlcolor = {blue},
}

\usepackage[table,xcdraw]{xcolor}
\usepackage{longtable}
\usepackage{booktabs}
\usepackage{makecell}
\usepackage{multirow}

\usepackage{multicol}
\usepackage[breakable]{tcolorbox}
\usepackage{changepage} 
\usepackage{enumitem}
\usepackage{float} 
\usepackage{adjustbox} 

\usepackage{fixme}
\fxsetup{status=draft} 
\fxsetup{theme=color}
\fxuselayouts{inline,marginclue}
\fxsetup{mode=multiuser}
\FXRegisterAuthor{chr}{achr}{CHR}
\FXRegisterAuthor{om}{aom}{OM}

\newcommand{\paper}{\cite{paper}}
\newcommand{\website}{\cite{website}}

\newcommand{\lorreview}{\cite{lor-review}}

\begin{document}
%
\title{Structured dataset documentation: a datasheet for CheXpert}


\makeatletter
\newcommand{\linebreakand}{%
  \end{@IEEEauthorhalign}
  \hfill\mbox{}\par
  \mbox{}\hfill\begin{@IEEEauthorhalign}
}
\makeatother

\author{\IEEEauthorblockN{Christian Garbin}
\IEEEauthorblockA{Department of Computer\\and Electrical Engineering\\ and Computer Science\\
Florida Atlantic University}
\and
\IEEEauthorblockN{Pranav Rajpurkar}
\IEEEauthorblockA{Department of Computer Science\\
Stanford University}
\and
\IEEEauthorblockN{Jeremy Irvin}
\IEEEauthorblockA{Department of Computer Science\\
Stanford University}
\linebreakand 
\IEEEauthorblockN{\\Matthew P. Lungren}
\IEEEauthorblockA{Department of Radiology\\
Stanford University}
\and
\IEEEauthorblockN{Oge Marques}
\IEEEauthorblockA{Department of Computer\\and Electrical Engineering\\ and Computer Science\\
Florida Atlantic University}
}


%


\maketitle

\begin{abstract}
\subfile{abstract.tex}
\end{abstract}


%
\IEEEpeerreviewmaketitle

\section{Introduction}

\subfile{introduction.tex}


\section{CheXpert Datasheet}

\subfile{chexpert-datasheet.tex}

\bibliographystyle{IEEEtran}
\bibliography{main}

\end{document}

%% file: abstract.tex
Billions of X-ray images are taken worldwide each year. Machine learning, and deep learning in particular, has shown potential to help radiologists triage and diagnose images. However, deep learning requires large datasets with reliable labels.

The CheXpert dataset was created with the participation of board-certified radiologists, resulting in the strong ground truth needed to train deep learning networks.

Following the structured format of Datasheets for Datasets, this paper expands on the original CheXpert paper and other sources to show the critical role played by radiologists in the creation of reliable labels and to describe the different aspects of the dataset composition in detail.

Such structured documentation intends to increase the awareness in the machine learning and medical communities of the strengths, applications, and evolution of CheXpert, thereby advancing the field of medical image analysis.

Another objective of this paper is to put forward this dataset datasheet as an example to the community of how to create detailed and structured descriptions of datasets. We believe that clearly documenting the creation process, the contents, and applications of datasets accelerates the creation of useful and reliable models.

%% file: introduction.tex
CheXpert (\textbf{Che}st e\textbf{Xpert}) is a ``public dataset for chest radiograph interpretation, consisting of 224,316 chest radiographs of 65,240 patients labeled for the presence of 14 observations as positive, negative, or uncertain.''~\cite{paper} Labels for the training set were extracted from the X-ray reports with a rule-based labeler that ``capture[s] uncertainties present in the reports by using an uncertainty label.''~\cite{paper} Labels for the validation and test sets were extracted by board-certified radiologists.

In this paper we present a dataset datasheet for CheXpert. Dataset datasheets~\cite{Gebru2018} are structured descriptions of datasets that document the ``motivation, composition, collection process, recommended uses, and so on. Datasheets for datasets facilitate communication between dataset creators and dataset consumers, and encourage the machine learning community to prioritize transparency and accountability''.

These sources were used to compile the CheXpert datasheet: the CheXpert paper, \href{https://arxiv.org/abs/1901.07031}{CheXpert: A Large Chest Radiograph Dataset with Uncertainty Labels and Expert Comparison}~\cite{paper}, \href{https://stanfordmlgroup.github.io/competitions/chexpert/}{The CheXpert Stanford website} \cite{website}, and \href{https://github.com/stanfordmlgroup/chexpert-labeler}{The CheXpert labeler GitHub repository} \cite{labeler}. The paper is the primary source for the datasheet. Quoted text without attribution should be assumed to be from this source.

%% file: chexpert-datasheet.tex
\begin{figure*}[!htb]
    \centering
    \includegraphics[width=0.9\textwidth]{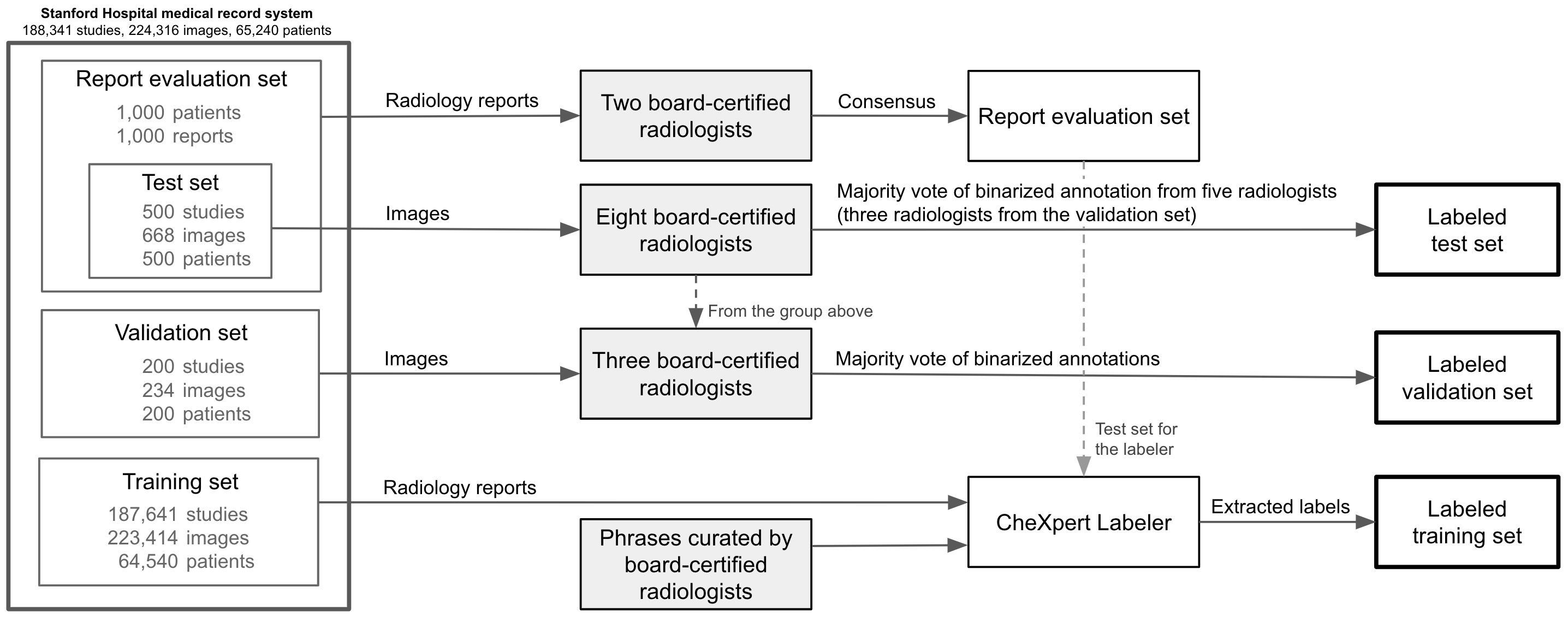}
    \caption[Sources for the training, validation, and test sets.]{Sources for the training, validation, and test sets, highlighting the participation of the board-certified radiologists to create a strong ground truth for each set.}
    \label{fig:sets-sources}
\end{figure*}

\definecolor{darkblue}{RGB}{46,25, 110}

\newcommand{\dssectionheader}[1]{%
   \noindent\framebox[\columnwidth]{%
      {\fontfamily{phv}\selectfont \textbf{\textcolor{darkblue}{#1}}}
   }
}

\newcommand{\dsquestion}[1]{%
    {\noindent \fontfamily{phv}\selectfont \textcolor{darkblue}{\textbf{#1}}}
}

\newcommand{\dsquestionex}[2]{%
    {\noindent \fontfamily{phv}\selectfont \textcolor{darkblue}{\textbf{#1} #2}}
}

\newcommand{\dsanswer}[1]{%
   {\noindent #1 \medskip}
}

\begin{singlespace}

\dssectionheader{Motivation}

\dsquestionex{For what purpose was the dataset created?}{Was there a specific task in mind? Was there a specific gap that needed to be filled? Please provide a description.}

\dsanswer{
``Chest radiography is the most common imaging examination globally, critical for screening, diagnosis, and management of many life threatening diseases. Automated chest radiograph interpretation at the level of practicing radiologists could provide substantial benefit in many medical settings, from improved workflow prioritization and clinical decision support to large-scale screening and global population health initiatives. For progress, there is a need for labeled datasets that (1) are large, (2) have strong reference standards, and (3) provide expert human performance metrics for comparison.''

``One of the main obstacles in the development of chest radiograph interpretation models has been the lack of datasets with strong radiologist-annotated groundtruth and expert scores against which researchers can compare their models. There are few chest radiographic imaging datasets that are publicly available, but none of them have test sets with strong ground truth or radiologist performances.''

Compared to existing chest X-ray datasets, CheXpert has the following advantages:

\begin{itemize}
    \item \textbf{Well-founded ground truth}: the validation and test sets were labeled by multiple board-certified radiologists, providing a strong ground truth to evaluate models.
    \item \textbf{Accurate label extraction}: a refined label extractor, capable of determining positive, negative, and uncertain mentions in radiology reports. The labeler uses ``[a] large list of phrases manually curated by multiple board-certified radiologists to match various ways observations are mentioned in the reports.'' The labeler was evaluated against labels manually extracted by board-certified radiologists.
\end{itemize}
}

\dsquestion{Who created this dataset (e.g., which team, research group) and on behalf of which entity (e.g., company, institution, organization)?}

\dsanswer{
CheXpert was created by a team of Stanford University's researchers from the Department of Computer Science, the Department of Medicine, and the Department of Radiology. The list of authors is available in \paper.
}

\dsquestionex{Who funded the creation of the dataset?}{If there is an associated grant, please provide the name of the grantor and the grant name and number.}

\dsanswer{
The work was supported by the Stanford center for artificial intelligence in medicine and imaging.
}

\dsquestion{Any other comments?}

\dsanswer{
No.
}

\bigskip
\dssectionheader{Composition}

\dsquestionex{What do the instances that comprise the dataset represent (e.g., documents, photos, people, countries)?}{ Are there multiple types of instances (e.g., movies, users, and ratings; people and interactions between them; nodes and edges)? Please provide a description.}

\dsanswer{
Each instance in the dataset is a frontal or lateral chest X-ray image, patient information, and labels for fourteen observations.
}

\dsquestion{How many instances are there in total (of each type, if appropriate)?}

\dsanswer{
The training set has 223,414 images from 64,540 patients, the validation set has 234 images from 200 patients, and the test set has 668 images from 500 patients. Each patient is present in only one of the sets.

Table \ref{tab:patient-studies-images-train-validate} shows the total number of patients, studies, and images in the training and validation sets. Table \ref{tab:patient-images-stats-distribution} shows the distribution of the number of images per patient. Each row in the table lists the count, percentages, and cumulative percentages for the number of patients and images. For example, it shows that there are 2,910 patients with five images (4.5\% of the total number of patients) and that patients with one to five images add up to 84.7\% of the total number of \textit{patients}. Similarly, it shows that the patients with five images account for 14,550 images (6.5\% of the total number of images) and that patients with one to five images add up to 50.1\% of the total number of \textit{images}.

\input{tables/patient-studies-images-train-validate}
\input{tables/patient-images-stats-distribution}

}





\dsquestionex{Does the dataset contain all possible instances or is it a sample (not necessarily random) of instances from a larger set?}{ If the dataset is a sample, then what is the larger set? Is the sample representative of the larger set (e.g., geographic coverage)? If so, please describe how this representativeness was validated/verified. If it is not representative of the larger set, please describe why not (e.g., to cover a more diverse range of instances, because instances were withheld or unavailable).}

\dsanswer{
The dataset contains studies from patients that visited the inpatient and outpatient centers of the Stanford Hospital between October 2002 and July 2017 and had a chest X-ray taken.

Studies that had been prospectively labeled as one of "normal", "known abnormality", and "emergent abnormality"  were included in the analysis.

The studies are ordered chronologically in the dataset.
}


\dsquestionex{What data does each instance consist of? “Raw” data (e.g., unprocessed text or images) or features?}{In either case, please provide a description.}

\dsanswer{
Each instance is a frontal or lateral X-ray image. There are two versions of each image:

\begin{itemize}
    \item Large: Original size from DICOM, grayscale downsampled to 256 levels. Images are stored in JPEG format.
    \item Small: Size downsampled to approximately 390 $\times$ 320 pixels (images size vary because they were downsampled to 320 pixels preserving ratio of dimensions), grayscale downsampled to 256 levels. Images are stored in JPEG format.
\end{itemize}

The large and small versions can be downloaded separately. The size of the large dataset is 440 GB, the size of the small one is 11 GB.

Images are grouped by patient and, within each patient, by study.

Each image is accompanied by a description:

\begin{itemize}
    \item The patient's biological sex: ``Female'', ``Male'', or ``Unknown''.
    \item The patient's age at the time the study was done, in number of complete years. Note that, although a rare occurrence, the same patient may have had studies done at different ages (see, for example, patient 11).
    \item Whether the image is frontal or lateral.
    \item For frontal images, whether the image is anteriorposterior (``AP''), or posterioranterior (``PA'').
    \item Fourteen columns with observations extracted from the image report (training set) or the images (validation set).
\end{itemize}
}

\dsquestionex{Is there a label or target associated with each instance?}{If so, please provide a description.}

\dsanswer{
\medskip
\textbf{Observations}

Each image has labels assigned to fourteen observations. Eleven of these observations are pathologies. The remainder observations indicate other findings in the report, or ``no finding''.

The eleven pathologies are as follows (the terms are based on the ``Fleischner Society: Glossary of Terms for Thoracic Imaging'' \cite{Hansell2008} whenever applicable).

\begin{itemize}
    \item Atelectasis
    \item Cardiomegaly
    \item Consolidation
    \item Edema
    \item Enlarged Cardiomediastinum
    \item Fracture (recent or healed)
    \item Lung Lesion
    \item Lung Opacity
    \item Pleural Effusion
    \item Pleural Other
    \item Pneumonia (``despite being a clinical diagnosis, [it] was included as a label in order to represent the images that suggested primary infection as the diagnosis'')
    \item Pneumothorax
\end{itemize}

In addition to the pathologies, each image has labels for the following observations.

\begin{itemize}
    \item No Finding: captures the absence of pathologies. The ``No Finding'' observation is assigned a positive label (1) if there is no pathology classified as positive or uncertain. The list of pathologies analyzed extends beyond the pathologies used for labels (list above). The list of pathologies is in \href{https://github.com/stanfordmlgroup/chexpert-labeler/blob/master/phrases/mention/no_finding.txt}{this file on GitHub}. If none of the pathologies in that list is mentioned in the medical report, the image is assigned the ``No Finding'' label. The only other label that may appear together with ``No Finding'' is ``Support Devices''.
    
    \item Support Devices: a support medical device, e.g. a tube, valve, pacemaker, among others, is present in the image. The list of devices extracted from the reports is available in \href{https://github.com/stanfordmlgroup/chexpert-labeler/blob/master/phrases/mention/support_devices.txt}{this file on GitHub}.
\end{itemize}

\medskip
\textbf{Label for each observation}

Labels were generated as follows for each observation.

\begin{itemize}
    \item Training set: a rule-based labeler extracted labels from the \textit{Impression} section of X-ray reports, using a large list of phrases manually curated by board-certified radiologists.
    \item Validation set: three board-certified radiologists annotated each of the studies separately, ``classifying each observation into one of present, uncertain likely, uncertain unlikely, and absent. Their annotations were binarized such that all present and uncertain likely cases are treated as positive and all absent and uncertain unlikely cases are treated as negative (i.e. there is no ``uncertain'' label in this set). The majority vote of these binarized annotations is used to define a strong ground truth''. The radiologists worked with the images, not with the reports.
    \item Test set: eight board-certified radiologists annotated the studies separately. The majority vote of five radiologists, three of them from the validation set and two selected randomly from the other five, is used as ground truth (the remainder three radiologists were used to evaluate the models in \paper). The same binarization process used in the validation set was used here (also resulting in no ``uncertain'' label in this set). The radiologists worked with the images, not with the reports.
\end{itemize}

In addition to the training, validation, and test sets, a ``report evaluation set'' was used to test the labeler. See details later in this section.

Figure \ref{fig:sets-sources} summarizes how the training, validation and test sets were created. The figure highlights where the board-certified radiologists participated, ensuring a strong ground truth for the validation and test sets, as well as for the set used to validate the CheXpert labeler.

The labels annotated by the radiologists followed a hierarchical structure. For example, both ``atelectasis'' and ``consolidation'' also result in the higher-level label ``lung opacity'' being assigned to the image.

Possible values for the labels are described in Table \ref{tab:label-definition}.

\begin{table}[h!]
\small
\centering
\caption{Label definitions}
\label{tab:label-definition}
\begin{tabular}{p{0.12\columnwidth}p{0.35\columnwidth}p{0.35\columnwidth}}
\toprule
\multirow{2}{*}{Label} & \multicolumn{2}{c}{How the label was assigned}                                                                                                                                                 \\ \cmidrule(l){2-3} 
                       & Training set                                                                                            & Validation and test sets                                                                 \\ \midrule
Positive               & The labeler found ``at least one mention that is positively classified in the {[}radiology{]} report''. & The majority of the radiologists classified as ``present'' or ``uncertain likely'' in the image.  \\ \midrule[0.2pt]
Negative               & The labeler found ``at least one negatively classified mention''.                                       & The majority of the radiologists classified as ``absent'' or ``uncertain unlikely'' in the image. \\ \midrule[0.2pt]
Uncertain              & The labeler found ``no positively classified mentions and at least one uncertain mention''.             & Not used in validation and test sets.                                                \\ \midrule[0.2pt]
No mention             & The labeler found ``no mention of an observation''.                                                     & Not used in the validation and test sets.                                            \\ \bottomrule
\end{tabular}
\end{table}

Note that the ``'uncertain' label can capture both the uncertainty of a radiologist in the diagnosis as well as ambiguity inherent in the report (``heart size is stable'').''

Labels are encoded as follows in the .csv files that accompany the dataset.

\begin{itemize}
    \item Positive: 1.0
    \item Negative: 0.0
    \item Uncertain: -1.0
    \item No mention: '' (the empty string)
\end{itemize}

Tables \ref{tab:label-frequency-training} and \ref{tab:label-frequency-validation} show the count and percentage of labels for each observation in the training and validation sets, respectively. Percentages add up to 100\% across rows. They do not add up to 100\% across columns because an image may have more than one observation with the same label (for example, an image may be positive for ``pleural effusion'' and ``lung opacity''). Also note that the validation set does not use the ``uncertain'' and ``no mention'' labels because ``present'' and ``uncertain likely'' cases are treated as positive and ``absent'' and ``uncertain unlikely'' cases are treated as negative, as explained in Table \ref{tab:label-definition}.

\input{tables/label-frequency-training}

\input{tables/label-frequency-validation}

Table \ref{tab:observation-coincidence} shows the coincidence of positive labels for each observation, i.e. how many times an observation was mentioned (labeled ``positive'') with another observation in the same report.

\input{tables/observation-coincidence}

\medskip
\textbf{Rule-based label extraction process for the training set}

The label extraction for the training set is a three-step process described in detail in \paper. A summarized description of the steps in the process is as follows.

\begin{itemize}
    \item Mention extraction: using a large list of phrases manually curated by radiologists, extract mentions from the \textit{Impression} section of the radiology reports.
    \item Mention classification: classify the extracted mentions as negative, uncertain, or positive. ``The mention classification stage is a 3-phase pipeline consisting of pre-negation uncertainty, negation, and post-negation uncertainty. Each phase consists of rules which are matched against the mention; if a match is found, then the mention is classified accordingly (as uncertain in the first or third phase, and as negative in the second phase). If a mention is not matched in any of the phases, it is classified as positive.''
    \item Mention aggregation: ``Observations with at least one mention that is positively classified in the report is assigned a positive (1) label. An observation is assigned an uncertain (u) label if it has no positively classified mentions and at least one uncertain mention, and a negative label if there is at least one negatively classified mention. We assign (blank) if there is no mention of an observation.''
\end{itemize}

Before being used to generate the labels in the training set, the labeler was evaluated against labels manually extracted from 1000 radiologist reports by two board-certified radiologists (this set of labels is referred to as the ``report evaluation set'' in Figure \ref{fig:sets-sources}). ``[W]ithout access to additional patient information, [the two radiologists] annotated the reports to label whether each observation was mentioned as confidently present (1), confidently absent (0), uncertainly present (u), or not mentioned (blank), after curating a list of labeling conventions to adhere to. After both radiologists independently labeled each of the 1000 reports, disagreements were resolved by consensus discussion. The resulting annotations serve as ground truth on the report evaluation set.'' The performance of the labeler against this ground truth is available in Table 2 in \paper.

The source code for the labeler is available in \href{https://github.com/stanfordmlgroup/chexpert-labeler}{this GitHub repository}.

Of interest for medical experts, these items are available in the GitHub repository:

\begin{itemize}
    \item \href{https://github.com/stanfordmlgroup/chexpert-labeler/tree/master/phrases/mention}{Phrases used to extract the observations from the reports}.
    \item \href{https://github.com/stanfordmlgroup/chexpert-labeler/blob/master/phrases/mention/no_finding.txt}{Phrases used to determine ``no finding''}. Phrases in this list are considered findings, i.e. a report must not have any of these terms to be labeled as ``no finding''. Note that the list is larger than the list of diseases labeled in the dataset.
    \item \href{https://github.com/stanfordmlgroup/chexpert-labeler/tree/master/patterns}{Negation and uncertainty patterns}.
\end{itemize}

\medskip
\textbf{Other uses and improvements of the labeler}

The CheXpert labeler was used to label the MIMIC-CXR dataset, comprised of 377,110 chest X-ray images \cite{Johnson2019}.

CheXbert is a BERT-based labeler developed after the CheXPert labeler, with the participation of some of the CheXpert team members \cite{Smit2020}. We recommend reviewing CheXbert to understand some of the limitations of the rule-based method used in CheXpert.

VisualCheXbert ``uses a biomedically-pretrained BERT model to directly map from a radiology report to the image labels, with a supervisory signal determined by a computer vision model trained to detect medical conditions from chest X-ray images'' \cite{Jain2021}. This method significantly improves the agreement between the labeler and radiologists labeling X-ray images.
}

\dsquestionex{Is any information missing from individual instances?}{If so, please provide a description, explaining why this information is missing (e.g., because it was unavailable). This does not include intentionally removed information, but might include, e.g., redacted text.}

\dsanswer{
All instances are complete.
}

\dsquestionex{Are relationships between individual instances made explicit (e.g., users’ movie ratings, social network links)?}{If so, please describe how these relationships are made explicit.}

\dsanswer{
Other than visiting the Stanford Hospital on those dates, there are no known relationships between the patients.
}

\dsquestionex{Are there recommended data splits (e.g., training, development/validation, testing)?}{If so, please provide a description of these splits, explaining the rationale behind them.}

\dsanswer{
The public dataset is split into test and validation sets, each in one folder. The list of training images is described in train.csv and the list of validation images is described in valid.csv (both files are delivered with the dataset). See tables \ref{tab:patient-studies-images-train-validate}, \ref{tab:label-frequency-training}, and \ref{tab:label-frequency-validation} for details of the split between the training and the validation sets.

A test set is used to score submissions to the leaderboard. This set is not publicly available.

Each patient is present in only one of the sets.

\medskip
\textbf{Training paradigms to handle uncertain observation} 

A significant innovation of the labeler used in CheXpert is the ability to label observations as ``uncertain''. Uncertain observations can be handled in different ways when training a classifier:

\begin{itemize}
    \item Ignore the uncertain labels. Because the uncertain label is quite prevalent in the training set, this may reduce the effective size of the set.
    \item Binary map the uncertain label to either positive or negative. Because the uncertainty label captures semantically useful information, this approach may degrade the classifier's performance.
    \item Self-train a classifier, using the uncertain label as ``unlabeled''.
    \item Treat uncertain as its own class and train a three-class classifier.
\end{itemize}

The paper \paper explores these approaches. Refer to the ``Model'' section in that paper for details.
}

\dsquestionex{Are there any errors, sources of noise, or redundancies in the dataset?}{If so, please provide a description.}

\dsanswer{
\textbf{Training set}

Extracting labels from reports with text-parsing tools is inherently imprecise. To reduce the errors in the extraction process, the following measures were taken to develop and test the labeler:

\begin{itemize}
    \item Development: the list of phrases used in the extraction phase ``was manually curated by multiple board-certified radiologists to match various ways observations are mentioned in the reports''. The list of phrases is available \href{https://github.com/stanfordmlgroup/chexpert-labeler/tree/master/phrases}{in these files on GitHub}. The list of negation and uncertainty patterns is available \href{https://github.com/stanfordmlgroup/chexpert-labeler/tree/master/patterns}{in these files on GitHub}.
    \item Evaluation: the labeler was evaluated against 1000 radiology reports that were manually processed by two board-certified radiologists. These reports were used as ground truth for the labeler. The F1 scores for the labeler on this reference set are available in Table 2 in \paper.
\end{itemize}

The labeler is described in detail in the section ``Label Extraction from Radiology Reports'' of \paper. The code is available in \href{https://github.com/stanfordmlgroup/chexpert-labeler}{this GitHub repository}.

\medskip
\textbf{Validation and test sets}

Labels in the validation and test sets were manually extracted from the images by a group of board-certified radiologists, using majority voting in case of disagreements. While this method cannot exclude errors, it is an accurate process to generate reliable labels from experts.
}


\dsquestionex{Is the dataset self-contained, or does it link to or otherwise rely on external resources (e.g., websites, tweets, other datasets)?}{If it links to or relies on external resources, a) are there guarantees that they will exist, and remain constant, over time; b) are there official archival versions of the complete dataset (i.e., including the external resources as they existed at the time the dataset was created); c) are there any restrictions (e.g., licenses, fees) associated with any of the external resources that might apply to a future user? Please provide descriptions of all external resources and any restrictions associated with them, as well as links or other access points, as appropriate.}

\dsanswer{
The dataset is self-contained.
}

\dsquestionex{Does the dataset contain data that might be considered confidential (e.g., data that is protected by legal privilege or by doctor-patient confidentiality, data that includes the content of individuals non-public communications)?}{If so, please provide a description.}

\dsanswer{
There is no private data in the dataset. Personally identifiable information has been removed from the data.
}

\dsquestionex{Does the dataset contain data that, if viewed directly, might be offensive, insulting, threatening, or might otherwise cause anxiety?}{If so, please describe why.}

\dsanswer{
No.
}

\dsquestionex{Does the dataset relate to people?}{If not, you may skip the remaining questions in this section.}

\dsanswer{
Yes. The dataset contains chest X-ray images from human patients.
}

\dsquestionex{Does the dataset identify any subpopulations (e.g., by age, gender)?}{If so, please describe how these subpopulations are identified and provide a description of their respective distributions within the dataset.}

\dsanswer{
The patient biological sex and age (in complete years) at the time the image was taken are available for each image.

Table \ref{tab:demographic-by-set-sex} shows the distribution of patients and images by sex in the training and validation sets. Table \ref{tab:demographic-by-set-sex-age-group} shows the distributions of ages (grouped by \href{https://www.ncbi.nlm.nih.gov/pmc/articles/PMC1794003/}{MeSH age groups}) across sex.

\input{tables/demographic-by-set-sex}

\input{tables/demographic-by-set-sex-age-group}

}

\dsquestionex{Is it possible to identify individuals (i.e., one or more natural persons), either directly or indirectly (i.e., in combination with other data) from the dataset?}{If so, please describe how.}

\dsanswer{
No.
}

\dsquestionex{Does the dataset contain data that might be considered sensitive in any way (e.g., data that reveals racial or ethnic origins, sexual orientations, religious beliefs, political opinions or union memberships, or locations; financial or health data; biometric or genetic data; forms of government identification, such as social security numbers; criminal history)?}{If so, please provide a description.}

\dsanswer{
No.
}

\dsquestion{Any other comments?}

\dsanswer{
No.
}

\bigskip
\dssectionheader{Collection Process}

\dsquestionex{How was the data associated with each instance acquired?}{Was the data directly observable (e.g., raw text, movie ratings), reported by subjects (e.g., survey responses), or indirectly inferred/derived from other data (e.g., part-of-speech tags, model-based guesses for age or language)? If data was reported by subjects or indirectly inferred/derived from other data, was the data validated/verified? If so, please describe how.}

\dsanswer{
Images and the associated reports were extracted from the Stanford Hospital DICOM and electronic medical records systems.
}

\dsquestionex{What mechanisms or procedures were used to collect the data (e.g., hardware apparatus or sensor, manual human curation, software program, software API)?}{How were these mechanisms or procedures validated?}

\dsanswer{
The images and associated reports were extracted from the hospital DICOM and PACS systems.

X-ray images come from a variety of X-ray devices used over the years at the hospital. The device used for each image is not available.
}

\dsquestion{If the dataset is a sample from a larger set, what was the sampling strategy (e.g., deterministic, probabilistic with specific sampling probabilities)?}

\dsanswer{
Images that had protected health information (PHI) of any type were excluded from the dataset.
}

\dsquestion{Who was involved in the data collection process (e.g., students, crowdworkers, contractors) and how were they compensated (e.g., how much were crowdworkers paid)?}

\dsanswer{
A team from Stanford's Computer Science, Medicine, and Radiology departments worked together on this project. The project was led through Stanford’s \href{https://stanfordmlgroup.github.io/programs/aihc-bootcamp/}{AI for Healthcare Bootcamp}.

None of the team members, students, and radiologists were compensated for this work, beyond their regular compensation for the position they held in their respective departments.
}


\dsquestionex{Over what timeframe was the data collected? Does this timeframe match the creation timeframe of the data associated with the instances (e.g., recent crawl of old news articles)?}{If not, please describe the timeframe in which the data associated with the instances was created.}

\dsanswer{
The images and records were collected by the hospital staff as part of their normal work protocols between October 2002 and July 2017.

The dataset was created in 2019, by analyzing those images and records.
}

\dsquestionex{Were any ethical review processes conducted (e.g., by an institutional review board)?}{If so, please provide a description of these review processes, including the outcomes, as well as a link or other access point to any supporting documentation.}

\dsanswer{
The Stanford Hospital Institutional Review Board (IRB) reviewed and approved the work.
}


\dsquestionex{Does the dataset relate to people?}{If not, you may skip the remaining questions in this section.}

\dsanswer{
Yes. The dataset contains chest X-ray images from human patients.
}

\dsquestion{Did you collect the data from the individuals in question directly, or obtain it via third parties or other sources (e.g., websites)?}

\dsanswer{
The data comes from the Stanford Hospital systems.
}

\dsquestionex{Were the individuals in question notified about the data collection?}{If so, please describe (or show with screenshots or other information) how notice was provided, and provide a link or other access point to, or otherwise reproduce, the exact language of the notification itself.}

\dsanswer{
Patients were not directly notified. The Stanford Hospital, being a research institution, notifies all patients that their information may be shared with researchers, subject to privacy laws and approval from an independent review process \cite{Hospital2013}. 
}

\dsquestionex{Did the individuals in question consent to the collection and use of their data?}{If so, please describe (or show with screenshots or other information) how consent was requested and provided, and provide a link or other access point to, or otherwise reproduce, the exact language to which the individuals consented.}

\dsanswer{
Individual patient consent is waived for de-identified data in compliance with institutional IRB and federal guidelines.
}

\dsquestionex{If consent was obtained, were the consenting individuals provided with a mechanism to revoke their consent in the future or for certain uses?}{If so, please provide a description, as well as a link or other access point to the mechanism (if appropriate).}

\dsanswer{
There is no mechanism in place to remove a patient from the dataset. Patients can request to be removed from future research projects through the Stanford Hospital privacy policies.
}

\dsquestionex{Has an analysis of the potential impact of the dataset and its use on data subjects (e.g., a data protection impact analysis) been conducted?}{If so, please provide a description of this analysis, including the outcomes, as well as a link or other access point to any supporting documentation.}

\dsanswer{
All instances in the dataset were anonymized. There is no personally identifiable data in the dataset.
}


\dsquestion{Any other comments?}

\dsanswer{
No.
}

\bigskip
\dssectionheader{Preprocessing/cleaning/labeling}

\dsquestionex{Was any preprocessing/cleaning/labeling of the data done (e.g., discretization or bucketing, tokenization, part-of-speech tagging, SIFT feature extraction, removal of instances, processing of missing values)?}{If so, please provide a description. If not, you may skip the remainder of the questions in this section.}

\dsanswer{
Images in the large and small datasets have their grayscale reduced to 256 levels. The small images have been downsized from the original size to approximately 390 $\times$ 320 pixels.
}

\dsquestionex{Was the “raw” data saved in addition to the preprocessed/cleaned/labeled data (e.g., to support unanticipated future uses)?}{If so, please provide a link or other access point to the “raw” data.}

\dsanswer{
The raw images and reports are available to the original authors in the Stanford Hospital electronic record systems. The data will not be made publicly available to preserve patients' privacy.
}

\dsquestionex{Is the software used to preprocess/clean/label the instances available?}{If so, please provide a link or other access point.}

\dsanswer{
The software to extract and preprocess the images is not publicly available.

The software to extract the labels is available in \href{https://github.com/stanfordmlgroup/chexpert-labeler}{this GitHub repository}.
}

\dsquestion{Any other comments?}

\dsanswer{
No.
}

\bigskip
\dssectionheader{Uses}

\dsquestionex{Has the dataset been used for any tasks already?}{If so, please provide a description.}

\dsanswer{
Several models to detect the labels in images were published together with the dataset in \paper. The ``models \ldots take as input a single-view chest radiograph and output the probability of each of the 14 observations. When more than one view is available, the models output the maximum probability of the observations across the views.''

The models explore different approaches to the ``uncertain'' label: ignore the label, binary mapping (map all uncertain instances to one in one model and zero in another model), self-training (treat as unlabeled instances), and treat the label as its own class. The results are available in \paper.

The dataset has also been used by multiple teams to submit entries to the leaderboard in \website. As of March 2021, there were 167 submissions and 6,430 dataset downloads with unique email addresses.

CheXpert is the primary source for CheXphoto, ``a dataset of smartphone photos and synthetic photographic transformations of chest x-rays'' \cite{Phillips2020}. The goal of this dataset is to advance the use of machine learning in clinical workflows that use photos of x-ray films, instead of digital x-ray images. This workflow, taking pictures of x-ray films and forwarding to other clinicians, is common in developing regions.
}

\dsquestionex{Is there a repository that links to any or all papers or systems that use the dataset?}{If so, please provide a link or other access point.}

\dsanswer{
A partial list of papers that use the dataset is available in the leaderboard in \website.
}

\dsquestion{What (other) tasks could the dataset be used for?}

\dsanswer{
The presence of the ``uncertain'' label in the dataset can be exploited in the training process in different ways. For example, in can be used a form of semi-supervised training, as explained in \paper.

The presence of frontal and lateral X-rays for the same patient can also be used for experiments in pathology identification.
}

\dsquestionex{Is there anything about the composition of the dataset or the way it was collected and preprocessed/cleaned/labeled that might impact future uses?}{For example, is there anything that a future user might need to know to avoid uses that could result in unfair treatment of individuals or groups (e.g., stereotyping, quality of service issues) or other undesirable harms (e.g., financial harms, legal risks) If so, please provide a description. Is there anything a future user could do to mitigate these undesirable harms?}

\dsanswer{
The dataset was created from images that come from one institution, the Stanford Hospital. It is known that models trained with images from one institution may perform worse with images from other institutions. Because the images come from one institution, they also come from a limited number of X-ray devices. It is known that the device used to capture images affects the model performance. \cite{Zech2018} \cite{Pooch2019} \cite{Yao2019}

There are indications in the literature that the sex and age of patients may affect the performance of models \cite{Larrazabal2020} \cite{Fonseca2020}. The distribution of sex in CheXpert images is generally well-balanced. However, there are few images of infants and adolescents. Models trained on this dataset may not perform well with pediatric images.

A small number of patients with large numbers of images reduce the dataset variability. The ``6 to 10 images'' row in Table \ref{tab:patient-images-stats-distribution} shows that approximately 85\% of the patients account for 50\% of the images, thus about half of the images in the dataset come from 15\% of the patients. This may reduce the effective size of the dataset, affecting some applications \lorreview.
}

\dsquestionex{Are there tasks for which the dataset should not be used?}{If so, please provide a description.}

\dsanswer{
According to the license terms, the dataset must be used only for non-clinical, non-commercial research. The dataset has not been reviewed or approved by the Food and Drug Administration (FDA).

The dataset is not suitable for diagnosing the pathologies. It can be used to determine the probability of the presence of the pathologies in research projects.
}

\dsquestion{Any other comments?}

\dsanswer{
No.
}

\bigskip
\dssectionheader{Distribution}

\dsquestionex{Will the dataset be distributed to third parties outside of the entity (e.g., company, institution, organization) on behalf of which the dataset was created?}{If so, please provide a description.}

\dsanswer{
The dataset is publicly available on \href{https://aimi.stanford.edu/research/public-datasets}{Stanford's Center for Artificial Intelligence in Medicine \& Imaging public datasets page}.
}

\dsquestionex{How will the dataset will be distributed (e.g., tarball on website, API, GitHub)}{Does the dataset have a digital object identifier (DOI)?}

\dsanswer{
The dataset is distributed as a compressed (.zip) file that contains all images and two comma-separated (.csv) files describing the training and the validation sets.

There are two versions of the dataset, each is a .zip file. One version has the downscaled version of the images (about 11 GB), the other has the original images (about 440 GB).
}

\dsquestion{When will the dataset be distributed?}

\dsanswer{
The dataset is currently available.
}

\dsquestionex{Will the dataset be distributed under a copyright or other intellectual property (IP) license, and/or under applicable terms of use (ToU)?}{If so, please describe this license and/or ToU, and provide a link or other access point to, or otherwise reproduce, any relevant licensing terms or ToU, as well as any fees associated with these restrictions.}

\dsanswer{
The license is available on the \href{https://stanfordmlgroup.github.io/competitions/chexpert/}{Stanford's Machine Learning GitHub} \website.

The main terms of the license are: for personal, non-commercial, non-clinical research use. Refer to the license for other terms of use.
}

\dsquestionex{Have any third parties imposed IP-based or other restrictions on the data associated with the instances?}{If so, please describe these restrictions, and provide a link or other access point to, or otherwise reproduce, any relevant licensing terms, as well as any fees associated with these restrictions.}

\dsanswer{
No.
}

\dsquestionex{Do any export controls or other regulatory restrictions apply to the dataset or to individual instances?}{If so, please describe these restrictions, and provide a link or other access point to, or otherwise reproduce, any supporting documentation.}

\dsanswer{
No.
}

\dsquestion{Any other comments?}

\dsanswer{
No.
}

\bigskip
\dssectionheader{Maintenance}

\dsquestion{Who will be supporting/hosting/maintaining the dataset?}

\dsanswer{
The Stanford Machine Learning Group.
}

\dsquestion{How can the owner/curator/manager of the dataset be contacted (e.g., email address)?}

\dsanswer{
Please refer to the \href{https://stanfordmlgroup.github.io/competitions/chexpert/}{CheXpert Stanford website}~\website.
}

\dsquestionex{Is there an erratum?}{If so, please provide a link or other access point.}

\dsanswer{
No.
}

\dsquestionex{Will the dataset be updated (e.g., to correct labeling errors, add new instances, delete instances)?}{If so, please describe how often, by whom, and how updates will be communicated to users (e.g., mailing list, GitHub)?}

\dsanswer{
There are no plans to update the dataset at this time.
}


\dsquestionex{If the dataset relates to people, are there applicable limits on the retention of the data associated with the instances (e.g., were individuals in question told that their data would be retained for a fixed period of time and then deleted)?}{If so, please describe these limits and explain how they will be enforced.}

\dsanswer{
The images and reports used to build the dataset are subject to retention policies of medical records in the Stanford Hospital.

There are no limits on the retention of the information in the dataset.
}

\dsquestionex{Will older versions of the dataset continue to be supported/hosted/maintained?}{If so, please describe how. If not, please describe how its obsolescence will be communicated to users.}

\dsanswer{
No.
}

\dsquestionex{If others want to extend/augment/build on/contribute to the dataset, is there a mechanism for them to do so?}{If so, please provide a description. Will these contributions be validated/verified? If so, please describe how. If not, why not? Is there a process for communicating/distributing these contributions to other users? If so, please provide a description.}

\dsanswer{
Contributions to the labeler can be submitted directly to \href{https://github.com/stanfordmlgroup/chexpert-labeler}{its GitHub site}.
}

\dsquestion{Any other comments?}

\dsanswer{
No.
}

\end{singlespace}

%% file: tables/patient-studies-images-train-validate.tex
\begin{table}[h!]
\centering
\caption{Number of patients, studies, and images}
\label{tab:patient-studies-images-train-validate}
\begin{tabular}{lrrr}
\toprule
{} & Patients & Studies &  Images \\
\midrule
Training   &   64,540 & 187,641 & 223,414 \\
Validation &      200 &     200 &     234 \\
\bottomrule
\end{tabular}
\end{table}

%% file: tables/patient-images-stats-distribution.tex
\begin{table*}[h!]
\small
\centering
\caption{Distribution of number of images per patient}
\label{tab:patient-images-stats-distribution}
\begin{adjustbox}{width = 0.75\textwidth}
\begin{tabular}{llrrrrrr}
\toprule
           &          & \multicolumn{3}{c}{Patients} & \multicolumn{3}{c}{Images} \\
           \cmidrule(lr){3-5}                  
           \cmidrule(lr){6-8}
           &          &    Count &    \% & Cum. \% &  Count &    \% & Cum. \% \\
\midrule
Training & 1 image &   22,581 & 35.0 &   35.0 & 22,581 & 10.1 &   10.1 \\
           & 2 images &   17,547 & 27.2 &   62.2 & 35,094 & 15.7 &   25.8 \\
           & 3 images &    6,965 & 10.8 &   73.0 & 20,895 &  9.4 &   35.2 \\
           & 4 images &    4,685 &  7.3 &   80.2 & 18,740 &  8.4 &   43.6 \\
           & 5 images &    2,910 &  4.5 &   84.7 & 14,550 &  6.5 &   50.1 \\
           & 6 to 10 images &    6,330 &  9.8 &   94.5 & 47,197 & 21.1 &   71.2 \\
           & 11 to 20 images &    2,665 &  4.1 &   98.7 & 37,636 & 16.8 &   88.0 \\
           & 21 to 30 images &      542 &  0.8 &   99.5 & 13,197 &  5.9 &   93.9 \\
           & More than 30 images &      315 &  0.5 &  100.0 & 13,524 &  6.1 &  100.0 \\
\midrule[0.2pt]
Validation & 1 image &      169 & 84.5 &   84.5 &    169 & 72.2 &   72.2 \\
           & 2 images &       28 & 14.0 &   98.5 &     56 & 23.9 &   96.2 \\
           & 3 images &        3 &  1.5 &  100.0 &      9 &  3.8 &  100.0 \\
\bottomrule
\end{tabular}
\end{adjustbox}
\end{table*}

%% file: tables/label-frequency-training.tex
\begin{table*}[h!]
\scriptsize
\centering
\caption{Frequency of labels in the training set images}
\label{tab:label-frequency-training}
\begin{adjustbox}{width = 0.9\textwidth}
\begin{tabular}{lrrrrrrrr}
\toprule
{} & Positive &    \% & Negative &    \% & Uncertain &    \% & No mention &    \% \\
\midrule
No Finding                 &   22,381 & 10.0 &        0 &  0.0 &         0 &  0.0 &    201,033 & 90.0 \\
Fracture                   &    9,040 &  4.0 &    2,512 &  1.1 &       642 &  0.3 &    211,220 & 94.5 \\
Support Devices            &  116,001 & 51.9 &    6,137 &  2.7 &     1,079 &  0.5 &    100,197 & 44.8 \\
Atelectasis                &   33,376 & 14.9 &    1,328 &  0.6 &    33,739 & 15.1 &    154,971 & 69.4 \\
Cardiomegaly               &   27,000 & 12.1 &   11,116 &  5.0 &     8,087 &  3.6 &    177,211 & 79.3 \\
\midrule[0.2pt]
Consolidation              &   14,783 &  6.6 &   28,097 & 12.6 &    27,742 & 12.4 &    152,792 & 68.4 \\
Edema                      &   52,246 & 23.4 &   20,726 &  9.3 &    12,984 &  5.8 &    137,458 & 61.5 \\
Enlarged Card. &   10,798 &  4.8 &   21,638 &  9.7 &    12,403 &  5.6 &    178,575 & 79.9 \\
Lung Lesion                &    9,186 &  4.1 &    1,270 &  0.6 &     1,488 &  0.7 &    211,470 & 94.7 \\
\midrule[0.2pt]
Lung Opacity               &  105,581 & 47.3 &    6,599 &  3.0 &     5,598 &  2.5 &    105,636 & 47.3 \\
Pleural Effusion           &   86,187 & 38.6 &   35,396 & 15.8 &    11,628 &  5.2 &     90,203 & 40.4 \\
Pleural Other              &    3,523 &  1.6 &      316 &  0.1 &     2,653 &  1.2 &    216,922 & 97.1 \\
Pneumonia                  &    6,039 &  2.7 &    2,799 &  1.3 &    18,770 &  8.4 &    195,806 & 87.6 \\
Pneumothorax               &   19,448 &  8.7 &   56,341 & 25.2 &     3,145 &  1.4 &    144,480 & 64.7 \\
\bottomrule
\end{tabular}
\end{adjustbox}
\end{table*}

%% file: tables/label-frequency-validation.tex
\begin{table}[h!]
\small
\centering
\caption{Frequency of labels in the validation set images}
\label{tab:label-frequency-validation}
\begin{tabular}{lrrrr}
\toprule
{} & Positive &    \% & Negative &     \% \\
\midrule
No Finding                 &       38 & 16.2 &      196 &  83.8 \\
Fracture                   &        0 &  0.0 &      234 & 100.0 \\
Support Devices            &      107 & 45.7 &      127 &  54.3 \\
Atelectasis                &       80 & 34.2 &      154 &  65.8 \\
Cardiomegaly               &       68 & 29.1 &      166 &  70.9 \\
\midrule[0.2pt]
Consolidation              &       33 & 14.1 &      201 &  85.9 \\
Edema                      &       45 & 19.2 &      189 &  80.8 \\
Enlarged Card. &      109 & 46.6 &      125 &  53.4 \\
Lung Lesion                &        1 &  0.4 &      233 &  99.6 \\
\midrule[0.2pt]
Lung Opacity               &      126 & 53.8 &      108 &  46.2 \\
Pleural Effusion           &       67 & 28.6 &      167 &  71.4 \\
Pleural Other              &        1 &  0.4 &      233 &  99.6 \\
Pneumonia                  &        8 &  3.4 &      226 &  96.6 \\
Pneumothorax               &        8 &  3.4 &      226 &  96.6 \\
\bottomrule
\end{tabular}
\end{table}

%% file: tables/observation-coincidence.tex
\begin{table*}[h!]
\centering
\caption{Coincidence of positive observations in the training set images}
\label{tab:observation-coincidence}
\begin{adjustbox}{width = 1\textwidth}
\begin{tabular}{rrrrrrrrrrrrrr}
\toprule
{} & \rotatebox{90}{No Finding} & \rotatebox{90}{Fracture} & \rotatebox{90}{Support Devices} & \rotatebox{90}{Atelectasis} & \rotatebox{90}{Cardiomegaly} & \rotatebox{90}{Consolidation} & \rotatebox{90}{Edema} & \rotatebox{90}{Enlarged Card.} & \rotatebox{90}{Lung Lesion} & \rotatebox{90}{Lung Opacity} & \rotatebox{90}{Pleural Effusion} & \rotatebox{90}{Pleural Other} & \rotatebox{90}{Pneumonia} \\
\midrule
Fracture                   &          0 &          &                 &             &              &               &        &                            &             &              &                  &               &           \\
Support Devices            &       8808 &     3628 &                 &             &              &               &        &                            &             &              &                  &               &           \\
Atelectasis                &          0 &     1281 &           20166 &             &              &               &        &                            &             &              &                  &               &           \\
Cardiomegaly               &          0 &      801 &           15539 &        3828 &              &               &        &                            &             &              &                  &               &           \\
\midrule[0.2pt]
Consolidation              &          0 &      401 &            7728 &        2003 &         1436 &               &        &                            &             &              &                  &               &           \\
Edema                      &          0 &     1009 &           33491 &        8844 &        11659 &          3154 &        &                            &             &              &                  &               &           \\
Enlarged Card. &          0 &      637 &            5614 &        1491 &         1901 &           585 &   2129 &                            &             &              &                  &               &           \\
Lung Lesion                &          0 &      343 &            3212 &        1025 &          623 &           748 &    866 &                        509 &             &              &                  &               &           \\
\midrule[0.2pt]
Lung Opacity               &          0 &     3611 &           61338 &       14146 &        13049 &          5574 &  27455 &                       5100 &        5316 &              &                  &               &           \\
Pleural Effusion           &          0 &     2456 &           52869 &       16290 &        11836 &          7403 &  26521 &                       3888 &        3265 &        52198 &                  &               &           \\
Pleural Other              &          0 &      296 &            1357 &         389 &          325 &           300 &    304 &                        193 &         310 &         1879 &              920 &               &           \\
Pneumonia                  &          0 &      141 &            1765 &         615 &          495 &          1014 &   1204 &                        150 &         454 &         4025 &             1733 &           123 &           \\
Pneumothorax               &          0 &     1050 &           11676 &        3131 &          777 &           739 &   1479 &                        773 &         866 &         9063 &             6691 &           253 &       146 \\
\bottomrule
\end{tabular}
\end{adjustbox}
\end{table*}

%% file: tables/demographic-by-set-sex.tex
\begin{table}[h!]
\small
\centering
\caption{Patients and images by sex}
\label{tab:demographic-by-set-sex}
\begin{tabular}{llrrrr}
\toprule
           &      & \multicolumn{2}{c}{Patients} & \multicolumn{2}{c}{Images} \\
           \cmidrule(lr){3-4}                  
           \cmidrule(lr){5-6}
           &      &    Count &    \% &   Count &    \% \\
\midrule
Training & Female &   28,729 & 44.5 &  90,778 & 40.6 \\
           & Male &   35,811 & 55.5 & 132,636 & 59.4 \\
\midrule[0.2pt]
Validation & Female &       94 & 47.0 &     106 & 45.3 \\
           & Male &      106 & 53.0 &     128 & 54.7 \\
\bottomrule
\end{tabular}
\end{table}

%% file: tables/demographic-by-set-sex-age-group.tex
\begin{table*}[h!]
\small
\centering
\caption{Patients, studies, and images by sex and age group}
\label{tab:demographic-by-set-sex-age-group}
\begin{adjustbox}{width = 0.9\textwidth}
\begin{tabular}{llrrrrrrrrr}
\toprule
           & {} & \multicolumn{3}{c}{Patients} & \multicolumn{3}{c}{Studies} & \multicolumn{3}{c}{Images} \\
           \cmidrule(lr){3-5}                  
           \cmidrule(lr){6-8}
           \cmidrule(lr){9-11}
           &  &   Female &   Male &    All &  Female &   Male &    All & Female &   Male &    All \\
\midrule
Training & (0-1) Infant &        3 &      0 &      3 &       3 &      0 &      3 &      3 &      0 &      3 \\
           & (13-18) Adolescent &      134 &    194 &    328 &     254 &    399 &    653 &    290 &    476 &    766 \\
           & (19-44) Adult &    5,666 &  7,458 & 13,124 &  14,395 & 20,055 & 34,450 & 16,972 & 24,724 & 41,696 \\
           & (45-64) Middle age &    9,607 & 13,309 & 22,916 &  27,026 & 42,033 & 69,059 & 31,653 & 51,097 & 82,750 \\
           & (65-79) Aged &    7,932 & 10,147 & 18,079 &  21,653 & 31,281 & 52,934 & 24,817 & 37,616 & 62,433 \\
           & (80+) Aged 80 &    6,118 &  5,598 & 11,716 &  14,846 & 15,696 & 30,542 & 17,043 & 18,723 & 35,766 \\
\midrule[0.2pt]
Validation & (13-18) Adolescent &        1 &      0 &      1 &       1 &      0 &      1 &      2 &      0 &      2 \\
           & (19-44) Adult &       19 &     14 &     33 &      19 &     14 &     33 &     22 &     16 &     38 \\
           & (45-64) Middle age &       35 &     40 &     75 &      35 &     40 &     75 &     39 &     49 &     88 \\
           & (65-79) Aged &       23 &     34 &     57 &      23 &     34 &     57 &     25 &     39 &     64 \\
           & (80+) Aged 80 &       16 &     18 &     34 &      16 &     18 &     34 &     18 &     24 &     42 \\
\bottomrule
\end{tabular}
\end{adjustbox}
\end{table*}